\documentclass[aip,reprint,twocolumn,amsmath,amssymb,showpacs]{revtex4-1}
\usepackage{graphicx,bm} \usepackage{color}

\begin{document}

\title{Broadband Transverse Electric Surface Wave in Silicene}

\author{M.~Shoufie Ukhtary} 
\email{shoufie@flex.phys.tohoku.ac.jp}
\author{Ahmad R. T. Nugraha}
\author{Eddwi H. Hasdeo}
\author{Riichiro Saito}

\affiliation{Department of Physics, Tohoku University, Sendai
  980-8578, Japan}

\begin{abstract}
  Transverse electric (TE) surface wave in silicine is theoretically
  investigated.  The TE surface wave in silicene is found to exhibit
  better characteristics compared with that in graphene, in terms of a
  broader frequency range and more confinement to the surface which
  originate from the buckled structure of silicene.  We found that
  even undoped silicene can support the TE surface wave.  We expect to
  obtain the similar characteristics of the TE surface wave in other
  two-dimensional materials that have slightly buckled honeycomb
  lattice.
\end{abstract}

\pacs{72.20.Pa,72.10.-d,73.50.Lw}
\date{\today}
\maketitle

Surface electromagnetic waves, or simply surface waves are
electromagnetic (EM) waves that propagate on the surface of a
material~\cite{hill2008light}.  Surface waves recently have attracted
a lot of interest, because of their capability to transport the EM
energy on the
surface~\cite{maier01,liu2006magnetic,oulton2008hybrid,vakil11,jablan09,hill2008light}.
There are two kinds of surface waves based on their polarizations; the
transverse magnetic (TM) and transverse electric (TE) surface
waves. In the case of TM surface wave, the component of magnetic field
is transverse to the propagation direction, while the electric field
has a component parallel to the propagation direction.  The TM surface
wave that also refers to a surface plasmon, can be seen as an electric
dipole wave on the surface of material due to spatial perturbation of
charge density~\cite{sun2014artificial,menabde15}.  On the other hand,
the TE surface wave has the component of electric field transverse to
the propagation direction while the magnetic field has a component
parallel to the direction of propagation. The TE surface wave can be
seen as a magnetic dipole wave on the surface of material due to the
self-sustained surface current
oscillation~\cite{sun2014artificial,menabde15}.

It is important to note that the radiation loss of magnetic dipole is
much smaller than that of electric dipole~\cite{jackson1999classical}.
Therefore, the TE surface wave can propagate longer than TM surface
wave~\cite{he14,liu2012manipulating}, which makes the TE surface wave
desirable for the transporting EM energy over long
distance~\cite{sun2014artificial,liu2012manipulating}.  However, the
TE surface wave cannot exist on the surface of an conventional bulk
metal because condition for generating the TE mode is limited which
means that the induced surface current is not available in the
conventional bulk
metal~\cite{maier01,sarid2010modern,hill2008light,sun2014artificial}.
Some efforts have been made for designing artificial materials so that
the TE surface wave can be generated, such as metamaterials and a
cluster of nanoparticles, which are generally complicated, hence
making them less viable and
accessible~\cite{sun2014artificial,ruppin2000surface,liu2012manipulating,liu2006magnetic}.

The difficulties of generating the TE surface wave can be alleviated
by using two-dimensional (2D) materials such like graphene, which is a
monolayer of carbon atoms arranged in honeycomb
lattice~\cite{novoselov05, mikhailov07}.  Mikhailov and Ziegler have
shown that, when the imaginary part of optical conductivity of 2D
material is negative (positive), the TE (TM) surface wave can
propagate on the surface of the 2D materials~\cite{mikhailov07}.  Due
to the presence of the Dirac cone in its electronic structure, the
imaginary part of optical conductivity of graphene can be negative at
a certain frequency range.  This is in contrast to usual 2D electron
gas systems, which have a positive imaginary part of optical
conductivity~\cite{mikhailov07,he14}.  This unusual property has also
enabled graphene to have the TE surface
wave~\cite{mikhailov07,he14,menabde15,jablan2011transverse}.  However,
it was predicted that the TE surface wave in doped graphene may only
exist for a narrow frequency range of $1.667
E_{\textrm{F}}<\hbar\omega<2
E_{\textrm{F}}$~\cite{mikhailov07,he14,menabde15}, where
$E_{\textrm{F}}$ is the Fermi energy.  Moreover, the TE surface wave
in graphene is less confined in the direction perpendicular to the
surface in comparison with the TM surface
wave~\cite{mikhailov07,he14}.

\begin{figure}[t]
  \centering\includegraphics[width=85mm]{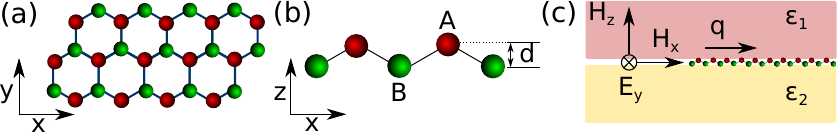}
\caption{(a) Honeycomb lattice of silicene. (b) Silicene lattice from
  side view. Sublattice A and B are separated vertically by
  $d$. Sublattice A (B) is depicted by red (green)
  atom. (c) Silicene is sandwiched between two dielectric
  media. TE surface wave propagates on the surface of silicene with
  wave vector $q$. The electric field $E_y$ is perpendicular to
  $q$. }
\label{fig1}
\end{figure}

In this letter, we propose that silicene is a better 2D material
rather than graphene to support the TE surface wave.  Silicene is a
monolayer of silicon atoms arranged in honeycomb lattice and the
stable structure of silicene is not purely planar, but slightly
buckled~\cite{stille12,tabert14,guzman07,ezawa2012topological}, i.e.,
the two sublattices are separated by vertical distance
$d=0.46~\mathrm{\AA}$ due to the $sp^{3}$-like
hybridization~\cite{liu2011quantum,ezawa2012topological}.  The
schematic structure of silicene can be seen in Figs.~\ref{fig1}(a) and
\ref{fig1}(b). The buckling of the atoms creates potential difference
between two sublattices when an external electric field is applied in
the direction perpendicular to the
surface~\cite{stille12,tabert14,guzman07,ezawa2012topological}. The
induced potential difference, along with the non-negligible spin orbit
(SO) coupling in silicene, will give a tunable energy
gap~\cite{liu2011low,stille12,tabert14,ezawa2012topological}.  We will
show that the tunable energy gap of silicene affects a unique optical
conductivity and the properties of TE surface wave which makes it a
key difference from graphene.

Suppose that a silicene layer, or generally any monolayer 2D material,
in the $x-y$ plane is sandwiched between two dielectric media with
dielectric constant $\varepsilon_1$ and $\varepsilon_2$ as shown in
Fig.~\ref{fig1}(c). The dispersion of the TE surface wave can be
obtained by employing the Maxwell equations with boundary conditions
of TE wave near the surface of the layer. Here we assume that the 2D
material is negligibly thin and it is characterized by its optical
conductivity $\sigma$ which will appear as a surface current density
in the boundary conditions for magnetic field as shown below. The TE
surface wave has an electric field $E_y$ in the $y$ direction and the
wave vector $q$ in the $x$ direction. There are two magnetic field
components $H_x$ and $H_z$ in TE surface wave as shown in
Fig.~\ref{fig1}(c).  Due to the confined nature of the surface wave,
the EM fields should decay in the direction perpendicular to the
surface ($z$). Thus, we can write the magnetic fields in the media $1$
and $2$ as $H_{x}^{(1)}(x,z)=A_{1}\exp(-\kappa_{1}z+iq x)$ and
$H_{x}^{(2)}(x,z)=A_{2}\exp(\kappa_{2}z+iq x)$, respectively.  The
electric field in the $k$-th medium ($k=1,2$) is obtained through
$E_{y}^{(k)}(x,z)=-i\omega\mu_0\int H_{x}^{(k)}(x,z) dz$.  The decay
constant $\kappa_k$ is given by
$\kappa_k=\sqrt{q^2-\varepsilon_k(\omega/c)^2}$.  The boundary
conditions at the surface are (i) $E_{y}^{(2)}-E_{y}^{(1)}=0$ and (ii)
$(H_{x}^{(2)}-H_{x}^{(1)})=J$, where $J=\sigma E_{y}$ is defined as
surface current density. Employing the boundary conditions and
assuming that the two dielectric media as vaccum
($\varepsilon_1=\varepsilon_2=1$, thus
$\kappa_1=\kappa_2=\sqrt{q^2-(\omega/c)^2}\equiv\kappa$), we obtain
the TE surface wave
dispersion~\cite{mikhailov07,jablan2011transverse},
\begin{align}
2 -
\frac{i\sigma(\omega)\omega\mu_0}{\kappa}=0~.
\label{eq:dispersion}
\end{align}
Since $\omega$ is a positive value, Eq.~(\ref{eq:dispersion}) requires
a negative value of Im~$\sigma$.  

\begin{figure}[t]
\centering\includegraphics[width=80mm]{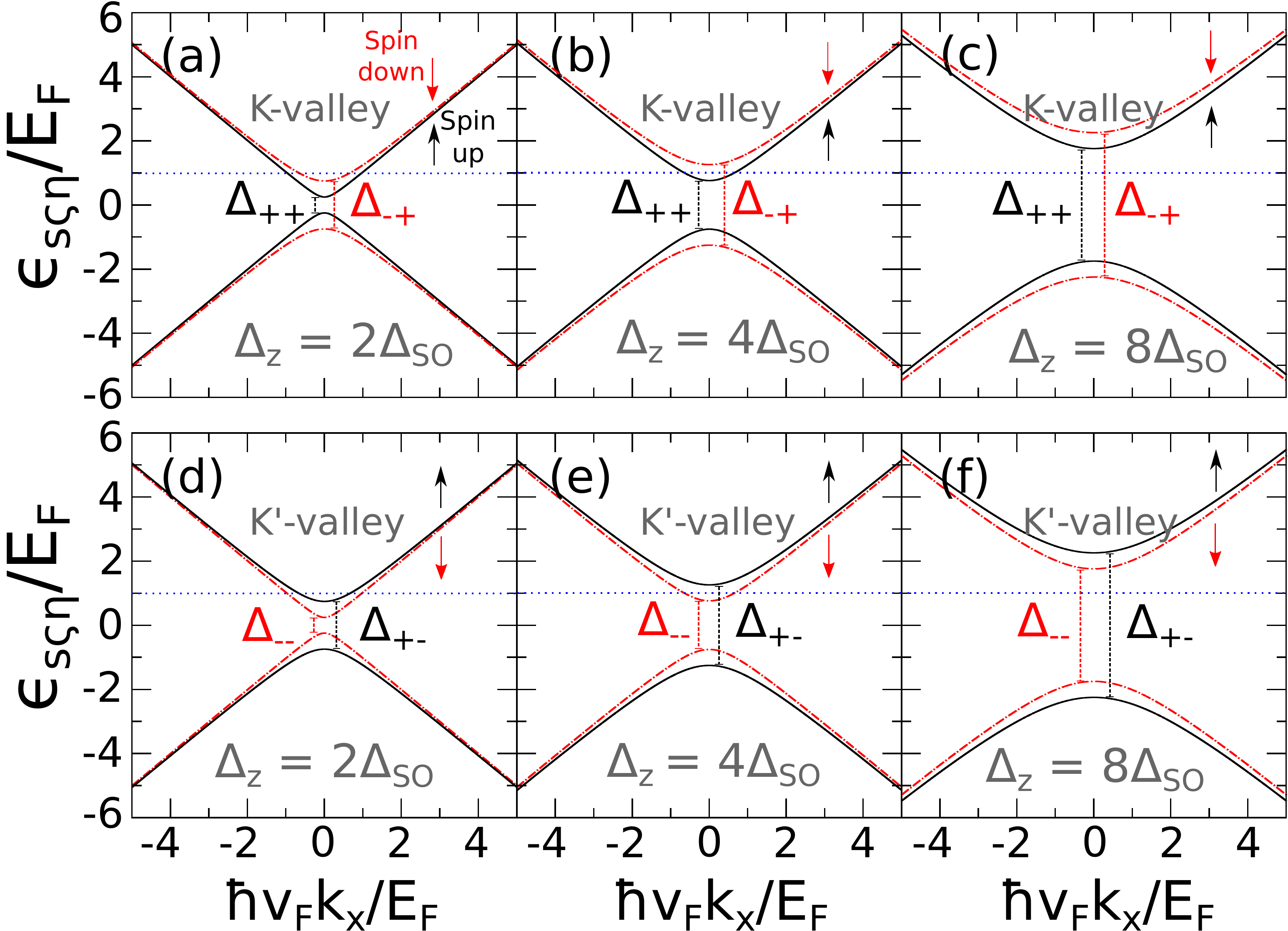}
\caption{\label{fig2} Electronic energy dispersions of silicene for
  [(a)-(c)] K and [(d)-(f)] K' valleys for several $\Delta_z$'s
  ($2\Delta_{\rm SO}$,$4\Delta_{\rm SO}$, and $8\Delta_{\rm SO}$).
  The solid (dash-dotted) lines correspond to spin up (down) electron
  dispersion. Positions of $E_{\textrm{F}}=7.8$~meV are indicated by
  the horizontal dotted lines.}
\end{figure}

Next, we derive the $\sigma (\omega)$ of silicene.  Similar to
graphene, the behavior of electrons at low-energy can be described by
the Dirac Hamiltonian near the $\textrm{K}$ and $\textrm{K}^\prime$
points (hexagonal corners of Brillouin
zone)~\cite{stille12,tabert14,guzman07,ezawa2012topological}.
However, we should consider the following two factors: (1) the SO
coupling in silicene is much larger than that of
graphene~\cite{liu2011low,stille12,tabert14}, and (2) the potential
difference between the sublattices A and B can be induced by an
external perpendicular electric field
$E_z$~\cite{stille12,tabert14,guzman07,ezawa2012topological}.  The
Hamiltonian of silicene can be written in the following matrix form,
\begin{equation}
  \label{eq:1}
  \begin{array}{ll}
    \widehat{H}_{\varsigma\eta} =
    \left [ 
      \begin{tabular}{cc}
        $-\frac{1}{2}\varsigma\eta\Delta_{\textrm{SO}}+\frac{1}{2}\Delta_z$
        & $\hbar v_{\textrm{F}} (\eta k_x-ik_y)$ \\
        $\hbar
        v_{\textrm{F}} (\eta k_x+ik_y)$ &
        $\frac{1}{2}\varsigma\eta\Delta_{\textrm{SO}}-\frac{1}{2}\Delta_z$
      \end{tabular}
    \right ]\\
  \end{array},
\end{equation}
where $v_{\rm F}$ is the Fermi velocity of electron and it is
$5.5\times 10^5 \textrm{m/s}$ for silicene~\cite{liu2011low}. The
Hamiltonian is spin and valley dependent, labeled by
$\varsigma=+1(-1)$ for spin up (spin down) and $\eta=+1(-1)$ for
$\textrm{K}$ ($\textrm{K}^\prime$)
valley. $\Delta_{\textrm{SO}}\approx 3.9~\textrm{meV}$ represents the
SO coupling for silicene~\cite{ezawa2012topological} and
$\Delta_z=\textrm{e}E_zd$ denotes the potential difference between
sublattices. $d=0.46~\mathrm{\AA}$ denotes the vertical distance
between the A and B atoms shown in Fig.~\ref{fig1}(b). The eigenvalues
of Eq.~(\ref{eq:1}) are expressed by
$\epsilon_{s\varsigma\eta}(k)=(-1)^{s+1}\epsilon_{\varsigma\eta}(k)$,
with $s$ is 1 and 2 for the conduction and valence band, respectively.
$\epsilon_{\varsigma\eta}(k)$ is the energy dispersion for electron
with $\varsigma$ spin and at $\eta$ valley, which is given by

\begin{align}
\epsilon_{\varsigma\eta}(k)=\sqrt{(\hbar
  v_{\textrm{F}}k)^2+\frac{1}{4}\Delta_{\varsigma\eta}^2}~,
\label{eq:2}
\end{align}
where $k=\sqrt{k_{x}^2+k_{y}^2}$ and $\Delta_{\varsigma\eta}(\Delta_z)
= \left|\Delta_z-\varsigma\eta\Delta_{\textrm{SO}}\right|$ denotes the
energy gap which is tunable by applying the $E_z$ up to
$2.6~\textrm{V}~\mathrm{\AA}^{-1}$ where the structure of silicene
becomes unstable~\cite{drummond2012electrically}. The energy gap is
defined as the energy separation from the top of the valence band to
the bottom of the conduction band with the same spin sign. There are
only two distinct values of $\Delta_{\varsigma\eta}(\Delta_z)$ for
four possible combination of $\Delta_{\varsigma\eta}$, since
$\Delta_{++}(\Delta_z)=\Delta_{--}(\Delta_z)$ and
$\Delta_{+-}(\Delta_z)=\Delta_{-+}(\Delta_z)$.

The optical conductivity $\sigma(\omega)$ of silicene can be obtained
by the Kubo formalism for current-current correlation
function~\cite{falkovsky2007space,bruus2004many}.  The electron scattering
is ignored here.  Following the derivation of graphene's
conductivity by Falkovsky and Varlamov, $\sigma_{\varsigma\eta} $ can
be expressed by~\cite{falkovsky2007space},
\begin{align}
\sigma_{\varsigma\eta} (\omega)=&
-\frac{ie^2}{4\omega\pi^2}\left\{\sum\limits_{s} \int d^2k [v_{ss}^x(k)]^2
\frac{df[\epsilon_{s\varsigma\eta}]}{d\epsilon_{s\varsigma\eta}}\right\}\nonumber\\
+&\frac{i\omega e^2}{2\pi^2}\hbar^2\Bigg\{\int d^2k
\frac{f[\epsilon_{1\varsigma\eta}(k)] - 
f[\epsilon_{2\varsigma\eta}(k)]}{\epsilon_{2\varsigma\eta}(k)
-\epsilon_{1\varsigma\eta}(k)}\nonumber\\
&\frac{v_{12}^x(k) v_{21}^x(k)}{\hbar^2\omega^2 - 
[\epsilon_{2\varsigma\eta}(k)-\epsilon_{1\varsigma\eta}(k)]^2}\Bigg\}~,
\label{eq:3}
\end{align}
where $f[\epsilon]$ is the Fermi distribution function and $v_{ss'}$
is the matrix element of velocity matrix
$\widehat{v}(k)=U^{-1}~(\partial
H_{\varsigma\eta}(\mathbf{k})/\hbar\partial \mathbf{k})~U$ in the $x$
direction, where $U$ is the unitary matrix which diagonalize
$H_{\varsigma\eta}$.  The $\widehat{v}(k)$ matrix is explicitly given
as follows:
\begin{widetext}
\begin{equation}
  \label{eq:4}
  \begin{array}{ll}
    \widehat{v}(k) =\frac{\hbar v_{\textrm{F}}^2 k}{\epsilon_{\varsigma\eta}}
      \Bigg[ &\begin{tabular}{cc}
        $\widehat{\bf{x}}\cos\theta+\widehat{\bf{y}}\sin\theta$
        &$-Z_{-}\{\widehat{\bf{x}}\eta(-\mathrm{\Gamma_{-}}+\mathrm{A_{-}})
        -\widehat{\bf{y}}i(\mathrm{B_{-}}-\mathrm{I_{-}})\}$
        \\ $Z_{+}\{\widehat{\bf{x}}\eta(\mathrm{\Gamma_{+}}+\mathrm{A_{+}})
        -\widehat{\bf{y}}i(\mathrm{B_{+}}+\mathrm{I_{+}})\}$
        &$-\widehat{\bf{x}}\cos\theta-\widehat{\bf{y}}\sin\theta$  
      \end{tabular}\Bigg]\\
  \end{array},
\end{equation}
\end{widetext}
where we define $\beta_\pm=\epsilon_{\varsigma\eta}\pm 1/2
\Delta_{\varsigma\eta}$, $Z_{\pm} =
\left(\beta_\pm/\beta_\mp\right)^{1/2}$, $\mathrm{\Gamma}_{\pm} =
\Delta_{\varsigma\eta}\cos\theta/\beta_\pm$, $\mathrm{A}_{\pm} =
i2\epsilon_{\varsigma\eta}\sin\theta/\beta_\pm$, $\mathrm{B}_{\pm} =
2\epsilon_{\varsigma\eta}\cos\theta/\beta_\pm$, and $\mathrm{I}_{\pm}
= i2\Delta_{\varsigma\eta}\sin\theta/\beta_\pm$.  Here $\theta$ is the
angle between $k_x$ and $k_y$, while $v_{nm}^x$ denotes the
$x$-component of the $n-m$ element of $\widehat{v}$ matrix. The first
(second) term in Eq.~(\ref{eq:3}) corresponds to the intraband
(interband) conductivity, which is later labeled as
$\sigma_{\varsigma\eta}^{\textrm{A}}$
($\sigma_{\varsigma\eta}^{\textrm{E}}$) .

By using Eqs.~(\ref{eq:2}) and~(\ref{eq:4}), we can calculate $\sigma$
in Eq.~(\ref{eq:3}) for silicene at
$T=0~\textrm{K}$~\cite{stille12,tabert14}.  Here, $\sigma$ is the
total conductivity for both spin and valley degrees of freedom. For
simplicity, we fix the $E_{\textrm{F}}=2\Delta_{\textrm{SO}}=7.8$~meV,
and vary the $\Delta_z$. Then, we get $\sigma$ as follows
\begin{align}
  \sigma(\omega,\Delta_z) =& \sum\limits_{\varsigma\eta}\left\{
    \sigma_{\varsigma\eta}^{\textrm{A}}(\omega,\Delta_z) +
    \sigma_{\varsigma\eta}^{\textrm{E}}(\omega,\Delta_z)\right\},\label{eq:sigtot}\\
  \sigma_{\varsigma\eta}^{\textrm{A}}(\omega,\Delta_z) =& i
  \frac{e^2}{16\hbar\pi} \frac{4E_{\textrm{F}}^2-
    \left[\Delta_{\varsigma\eta}(\Delta_z)\right]^2}{E_{\textrm{F}}\hbar\omega}
  \Theta\left[2E_{\textrm{F}}-\Delta_{\varsigma\eta}(\Delta_z)
  \right] \label{eq:intra}\\
  \sigma_{\varsigma\eta}^{\textrm{E}}(\omega,\Delta_z) =&
  \frac{e^2}{16\hbar}
  \left\{1+\Bigg(\frac{\Delta_{\varsigma\eta}(\Delta_z)}{\hbar\omega}
  \Bigg)^2\right\}
  \Theta\left[\hbar\omega-g(\Delta_z)\right]\nonumber\\ 
  &-i\frac{e^2}{16\hbar\pi}
  \left\{1+\Bigg(\frac{\Delta_{\varsigma\eta}(\Delta_z)}
                {\hbar\omega}\Bigg)^2\right\} \nonumber\\
                & \times\ln\left|\frac{\hbar\omega+ g(\Delta_z)
                }{\hbar\omega-g(\Delta_z)}\right|+
                i\frac{e^2\left[\Delta_{\varsigma\eta}(\Delta_z)\right]^2}
                {8\hbar^2\pi\omega g(\Delta_z) }\quad,
\label{eq:inter}
\end{align} 
where $\Theta(x)$ is the Heaviside function and
$g(\Delta_z)=\textrm{max}[2E_{\textrm{F}},\Delta_{\varsigma\eta}(\Delta_z)]$. If
we set $\Delta_{\varsigma\eta}=0$, we get the optical conductivity of
graphene~\cite{mikhailov07,he14}.

In Fig.~\ref{fig2}, we plot the electron energy dispersions for K and
K' valleys based on Eq.~(\ref{eq:2}) for several $\Delta_z$'s.  In
varying $\Delta_z$, we choose three cases for both the K and K'
valleys depending on the position of $E_{\textrm{F}}$ relative to the
energy gap, which are shown in Fig.~\ref{fig2}.  The first case is
$\Delta_z=2\Delta_{\textrm{SO}}$ , in which $E_{\textrm{F}}$ is higher
than bottoms of the two conduction bands for spin up and spin down
($E_{\textrm{F}}>\Delta_{++/--}$ and $\Delta_{-+/+-}$)
[Figs.~\ref{fig2}(a) and 2(d)].  The second case is
$\Delta_z=4\Delta_{\textrm{SO}}$, in which $E_{\textrm{F}}$ lies
between two bottoms of the conduction bands
($\Delta_{++/--}<E_{\textrm{F}}<\Delta_{-+/+-}$) [Figs.~\ref{fig2}(b)
  and 2(e)] and the third case is $\Delta_z=8\Delta_{\textrm{SO}}$, in
which $E_{\textrm{F}}$ exists in energy gaps [Figs.~\ref{fig2}(c) and
  2(f)].

\begin{figure}[t]
\centering\includegraphics[width=7cm]{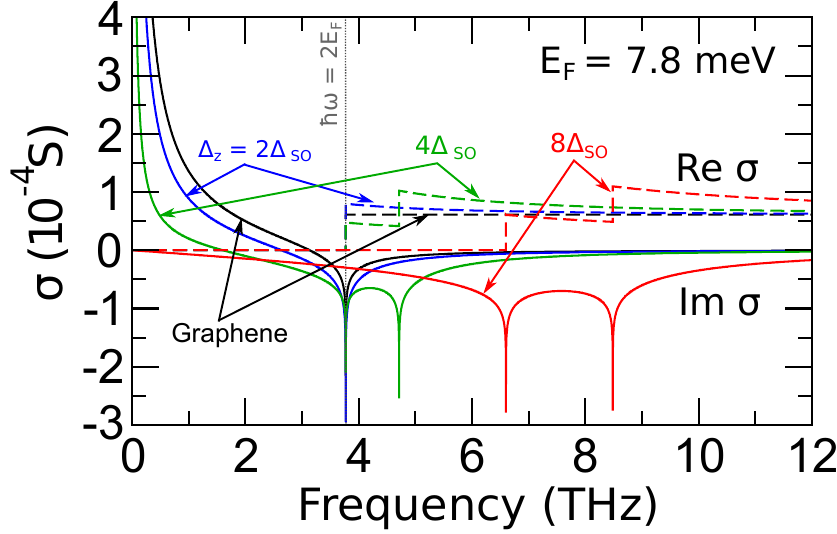}
\caption{\label{fig3} Optical conductivity $(\sigma)$ of silicene for
  three different $\Delta_z$ values and compared with that of
  graphene.  The solid lines represent the imaginary part of $\sigma$
  and the dashed lines represent the real part of $\sigma$.  Position
  of $\hbar\omega=2E_{\textrm{F}}$ is pointed at $f=3.77$~THz.}
\end{figure}

In Fig.~\ref{fig3}, we plot the optical conductivity $\sigma$ of
silicene as a function of frequency where the solid and dashed lines
are the imaginary and real parts of $\sigma$, respectively, for the
three $\Delta_z$. We also plot the $\sigma$ of graphene in black lines
for a comparison. The logarithmic singularities in Im~$\sigma$ in
Eq.~\eqref{eq:inter} correspond to the lowest excitation energies for
interband transitions of electrons between energy bands having the
same spin directions and the same valleys. Im~$\sigma$ is singular for
any frequency which satisfies condition $\hbar\omega=g(\Delta_z)$.
Since there are two distinct values of
$\Delta_{\varsigma\eta}(\Delta_z)$, there are two possible singularity
points, $\omega_1=2E_{\textrm{F}}/\hbar$ and
$\omega_2=\Delta_{-+}/\hbar$ if
$\Delta_{++}/2<E_{\textrm{F}}<\Delta_{-+}/2$
[$\Delta_z=4\Delta_{\textrm{SO}}$] or $\omega_1=\Delta_{++}/\hbar$ and
$\omega_2=\Delta_{-+}/\hbar$ if $E_{\textrm{F}}<\Delta_{++}/2$ and
$\Delta_{-+}/2$ [$\Delta_z=8\Delta_{\textrm{SO}}$]. When
$E_{\textrm{F}}>\Delta_{-+}/2$ and $\Delta_{++}/2$ there is only one
singularity point at $\omega=2E_{\textrm{F}}/\hbar$. Since $\sigma$ of
silicene depends on $\Delta_z$, $\sigma$ of silicence can be tuned not
only by $E_{\textrm{F}}$ but also by $E_z$. As mentioned in
Eq.~\eqref{eq:dispersion}, the negative value of Im~$\sigma$
correspond to the condition for TE surface wave. The TE surface wave
cannot exist for the region that Im $\sigma>0$. In the following
discussion, we call the frequency range of Im $\sigma<0$ as the TE
frequency range. Furthermore we focus only on the frequency range
where Re~$\sigma=0$ in which the TE surface wave is not
damped~\cite{mikhailov07}. For graphene ($\Delta_{\varsigma\eta}=0$),
the TE frequency range is fixed at
$1.667E_{\textrm{F}}<\hbar\omega<2E_{\textrm{F}}$
($3.14<f<3.77~\textrm{THz}$), which reproduces the previous
results~\cite{mikhailov07,he14}.

In general, the TE frequency range in silicene is wider than that in
graphene for the same $E_{\textrm{F}}$ and it is tunable by $\Delta_z$
as shown in Fig.~\ref{fig3}.  For example, for
$\Delta_z=2\Delta_{\textrm{SO}}~(E_z=16.96~\textrm{mV}\mathrm{\AA}^{-1})$,
the TE frequency range lies within
$1.4E_{\textrm{F}}<\hbar\omega<2E_{\textrm{F}}$~($ 2.64 < f
<3.77~\textrm{THz}$).  By increasing $\Delta_z$,
$\Delta_{\varsigma\eta}$ increases.  From
Eq.~(\ref{eq:intra})-(\ref{eq:inter}) we know that increasing
$\Delta_{\varsigma\eta}$ not only makes Im
$\sigma_{\varsigma\eta}^{\textrm{E}}$ more negative, but also reduces
Im $\sigma_{\varsigma\eta}^{\textrm{A}}$ whose value is always
positive~\cite{mikhailov07}.  Altogether, Im $\sigma$ decreases, hence
the TE frequency range becomes wider when we increase $\Delta_z$.  The
Im $\sigma_{\varsigma\eta}^{\textrm{A}}$ can be suppressed when
$\Delta_{\varsigma\eta}>4\Delta_{\textrm{SO}}$, or the Fermi level is
located in $\Delta_{\varsigma\eta}$ [Figs.~\ref{fig2}(b)--(c)]. This
occurs for
$\Delta_z=4\Delta_{\textrm{SO}}~(E_z=33.92~\textrm{mV}\mathrm{\AA}^{-1})$
and
$\Delta_z=8\Delta_{\textrm{SO}}~(E_z=67.84~\textrm{mV}\mathrm{\AA}^{-1})$
(see Figs.~\ref{fig2}(b) and (c) respectively). For
$\Delta_z=4\Delta_{\textrm{SO}}$, only Im $\sigma_{-+}^{\textrm{A}}$
and Im $\sigma_{+-}^{\textrm{A}}$ are suppressed, therefore we still
have Im $\sigma>0$ at certain frequency and Re $\sigma\neq0$ for
$\hbar\omega\geq2E_{\textrm{F}}$ ($f\geq3.77~\textrm{THz}$, see
Eqs.~(\ref{eq:intra})\ and (\ref{eq:inter})). Hence, the TE frequency
range becomes $1.61<f<3.77~\textrm{THz}$.  But in the case of
$\Delta_z=8\Delta_{\textrm{SO}}$, all Im
$\sigma_{\varsigma\eta}^{\textrm{A}}$ vanish and Im $\sigma$ has
negative value at all frequency. Re $\sigma\neq0$ for
$\hbar\omega\geq\Delta_{++}$ ($f\geq6.60~\textrm{THz}$). Therefore,
the TE frequency range becomes $0<f<6.60~\textrm{THz}$. Re $\sigma$
appears at higher frequency than that for
$\Delta_z=4\Delta_{\textrm{SO}}$, because the Fermi level exists in
all of the energy gaps, in which we need a higher excitation energy
for interband transition.

\begin{figure}[t]
\centering\includegraphics[width=7cm]{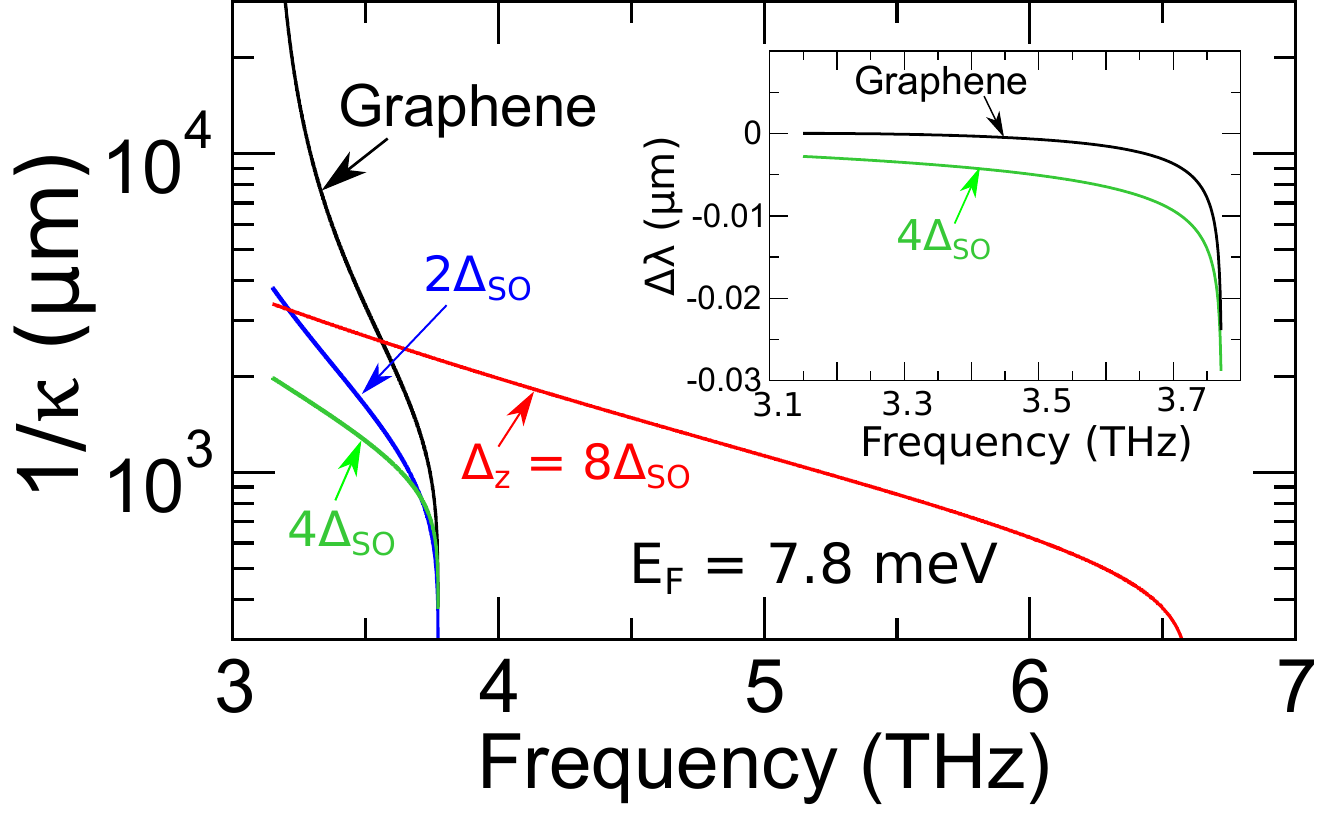}
\caption{Confinement length $1/\mathrm{\kappa}$ for silicene and
  graphene as a function of frequency. Inset : $\Delta\lambda$ as a function of 
frequency for graphene and silicene.}
\label{fig4}
\end{figure}

Another interesting finding is that the undoped silicene
($E_{\textrm{F}}=0$) may also support TE surface wave. From
Eqs.~(\ref{eq:sigtot})--(\ref{eq:inter}), for $E_{\textrm{F}}=0$ we
get Im $\sigma$:
\begin{align}
  \textrm{Im}~\sigma(\omega,\Delta_z)&
  =-\frac{e^2}{16\hbar\pi}\sum\limits_{\varsigma\eta}
  \Bigg\{\left[1+\left(\frac{\Delta_{\varsigma\eta}(\Delta_z)}
      {\hbar\omega}\right)^2\right]\nonumber\\
  &\times\ln\left|\frac{\hbar\omega+\Delta_{\varsigma\eta}(\Delta_z)}
    {\hbar\omega-\Delta_{\varsigma\eta}(\Delta_z)}
  \right|-\frac{2\Delta_{\varsigma\eta}(\Delta_z)}{\hbar\omega}\Bigg\}\quad.
\label{eq:undoped}
\end{align}
The TE frequency range lies within $0<\hbar\omega<\Delta_{++/--}$. It
is noted that $\textrm{Im}~\sigma(\omega)$ vanishes at
$E_{\textrm{F}}=0$ in graphene~\cite{mikhailov07}, hence the TE
surface wave does not exist for undoped graphene.

From Eq.~(\ref{eq:dispersion}), we can define a confinement length of
TE surface wave $1/\kappa$, as follows 
\begin{align}
\frac{1}{\kappa}=\frac{2}{i\omega\sigma(\omega,\Delta_z)\mu_0}\quad.
\label{eq:decay}
\end{align}
A smaller value of $1/\kappa$ corresponds to better confinement. In
Fig.~\ref{fig4}, we plot $1/\kappa$ of the TE surface wave in graphene
and silicene for comparison. The plot starts at
$f=3.15~\textrm{THz}~(\hbar\omega=1.667E_{\textrm{F}})$, which is the
lower bound of TE frequency range in graphene. We can see that the TE
surface wave in silicene is much more confined than in graphene and
tunable by $\Delta_z$. For example, at
$f=3.25~\textrm{THz}~(\hbar\omega=1.725E_{\textrm{F}})$, in case of
graphene, $1/\mathrm{\kappa}=13994~\mathrm{\mu m}$, while in case of
silicene, $1/\mathrm{\kappa}=2906.2~\mathrm{\mu m}$ for
$\Delta_z=2\Delta_{\textrm{SO}}$,
$1/\mathrm{\kappa}=1747.7~\mathrm{\mu m}$ for
$\Delta_z=4\Delta_{\textrm{SO}}$, and
$1/\mathrm{\kappa}=3146.7~\mathrm{\mu m}$ for
$\Delta_z=8\Delta_{\textrm{SO}}$. In the case of
$\Delta_z=8\Delta_{\textrm{SO}}$, we might get a larger
$1/\kappa$. This is because $\textrm{Im}~\sigma$ is singular at higher
frequency, which makes $1/\kappa$ for $\Delta_z =
8\Delta_{\textrm{SO}}$ slowly diverge.

By solving Eq.~(\ref{eq:dispersion}) for $\lambda=2\pi/q$, we can
define the difference between the wavelength of TE surface wave
$\lambda$ and the wavelength of freely propagating EM wave in vacuum
$\lambda_0=2\pi c/\omega$ as $\Delta\lambda=\lambda-\lambda_0$ .  In
the inset of Fig.~\ref{fig4} we plot $\Delta\lambda$ as a function of
frequency for graphene and silicene with
$\Delta_z=4\Delta_{\textrm{SO}}$.  We can see that $\Delta\lambda$ is
sufficiently small which means that $\lambda$ is almost the same as
$\lambda_0$ (3 THz corresponds to $\lambda_0=100\mathrm{\mu}$m).
However, $\Delta\lambda$ for silicene is more negative compared with
that for graphene, which is almost zero. Negative $\Delta\lambda$
means that there is shrinkage of the wavelength of TE surface wave
which is the preferable feature of surface wave since more information
can be compressed in the wave. From the inset of Fig.~\ref{fig4}, we
can see more shrinkage of the wavelength in silicene compared with
that in graphene.

In conclusion, silicene is theoretically proved to be a versatile
platform for utilizing TE surface wave.  We have shown that silicene
supports the TE surface wave propagation and it exhibits more
preferable surface wave properties compared with those of graphene,
such as the tunable broadband frequency and smaller confinement
length.  The TE surface wave in silicene is tunable by the Fermi
energy as well as by the external electric field.  These
characteristics originate from the two-dimensional buckled honeycomb
structure.

M.S.U. and E.H.H. are supported by the MEXT scholarship. A.R.T.N.
acknowledges the Leading Graduate School Program in Tohoku University.
R.S. acknowledges MEXT (Japan) Grants No. 25107005 and No. 25286005.


\end{document}